\documentclass[fleqn,usenatbib]{mnras}

\usepackage{newtxtext,newtxmath}

\usepackage[T1]{fontenc}

\DeclareRobustCommand{\VAN}[3]{#2}
\let\VANthebibliography\thebibliography
\def\thebibliography{\DeclareRobustCommand{\VAN}[3]{##3}\VANthebibliography}

\usepackage{graphicx}	
\usepackage{amsmath}	

\newcommand{\rc}{\color{black}}
\newcommand{\tworc}{\color{black}}

\title[]{Using helium 10830~\AA~ transits to constrain planetary magnetic fields}

\author[Schreyer et al.]{
Ethan Schreyer,$^{1}$\thanks{E-mail:ethan.schreyer16@imperial.ac.uk}
James E. Owen,$^{1}$
Jessica J. Spake,$^{2}$
Zahra Bahroloom$^{1}$ and Simone Di Giampasquale$^{1}$
\\
$^{1}$Astrophysics Group, Department of Physics, Imperial College London, Prince Consort Rd, London SW7 2AZ, UK\\
$^{2}$Division of Geological and Planetary Sciences, California Institute of Technology, Pasadena, CA 91125, USA
}

\date{Accepted XXX. Received YYY; in original form ZZZ}

\pubyear{2015}

\begin{document}
\label{firstpage}
\pagerange{\pageref{firstpage}--\pageref{lastpage}}
\maketitle

\begin{abstract}
Planetary magnetic fields can affect the predicted mass loss rate for close-in planets that experience large amounts of UV irradiation. In this work, we present a method to detect the magnetic fields of close-in exoplanets undergoing atmospheric escape using transit spectroscopy at the 10830 \AA~ line of helium. Motivated by previous work on hydrodynamic and magneto-hydrodynamic photoevaporation, we suggest that planets with magnetic fields that are too weak to control the outflow's topology lead to blue-shifted transits due to day-to-night-side flows. In contrast, strong magnetic fields prevent this day-to-night flow, as the gas is forced to follow the magnetic field's roughly dipolar topology. We post-process existing 2D photoevaporation simulations, computing synthetic transit profiles in helium to test this concept. As expected, we find that hydrodynamically dominated outflows lead to blue-shifted transits on the order of the sound speed of the gas. Strong surface magnetic fields lead to unshifted or slightly red-shifted transit profiles. High-resolution observations can distinguish between these profiles; however, eccentricity uncertainties generally mean that we cannot conclusively say velocity shifts are due to the outflow for individual planets. The majority of helium observations are blue-shifted, which could be a tentative indication that close-in planets generally have surface dipole magnetic field strengths $\lesssim 0.3$~Gauss. More 3D hydrodynamic and magneto-hydrodynamic simulations are needed to confirm this conclusion robustly.           
\end{abstract}




\section{Introduction}

Planets that reside close to their host stars receive significant amounts of high-energy flux. Photoionization of their upper atmospheres results in heating that can drive mass loss in the form of a hydrodynamic wind \citep{Lammer2003, GarciaMuniz2007, Murray-Clay2009, Owen2012, Kubyshkina2018}.
For small, low-mass planets ($1-4 \, {\rm R}_{\oplus}$, $\lesssim 20$~M$_\oplus$), photoevaporation can completely strip their primordial hydrogen/helium envelopes, leaving behind a ``stripped core'' with a  bulk composition similar to Earth's \citep{Owen2013, Lopez2013}. This photoevaporation-driven evolution has been used to explain the lack of short period Neptune sized planets \citep[e.g.][]{Szabo2011, Lundkvist2016, Owen2018}, and the bimodal radius distribution of small planets \citep[]{Fulton2017,Owen2017,VanEylen2018,Rogers2023}. However, other interpretations have been proposed to explain the demographics of close-in, low-mass exoplanets. For example, other mass loss mechanisms such as core-powered mass-loss \citep[e.g,][]{Ginzburg2018, Gupta2019} or impact-powered mass-loss \citep[e.g.][]{Wyatt2020}; late time gas accretion in the formation phase \citep[e.g,][]{Lee2021, Lee2022}, or separate populations of terrestrial planets and water worlds \citep[e.g,][]{Zeng2019, Luque2022}. 

Given this debate, we need to test photoevaporation models using direct observations of atmospheric escape. An unquantified uncertainty when testing these models is the role that magnetic fields play in controlling the outflow and the mass-loss \citep[e.g.][]{Owen2019b}. The photoevaporation models used to match the observed location of the radius valley have not included magnetic fields \citep[e.g,][]{Owen2017, Wu2019, Rogers2021a, Rogers2023b,Rogers2023}. Photoevaporative outflows are highly ionized, so they should be influenced by any planetary magnetic fields. The degree to which the planetary magnetic field controls the outflow is determined by the ratio of magnetic pressure to thermal/ram pressure of the outflow \citep[e.g.][]{Adams2011}. If the magnetic pressure dominates near the planetary surface, then the gas is constrained to the topology imposed by the planet's magnetic field. For this ``magnetically controlled'' case, analytic and numerical studies have demonstrated that in a region near the equator, the magnetic field lines are closed, and the gas is trapped in a ``dead-zone'' \citep[e.g,][]{Trammell2011, Adams2011, Trammell2014, Owen2014, Khodachenko2015, Matsakos2015, Arakcheev2017}. Hence, the gas can only escape near the poles, where the gas pressure can open magnetic field lines and drive a transonic wind. Alternatively, if the thermal/ram pressure dominates everywhere, the gas will open all field lines (dragging the magnetic field with it). In this ``hydrodynamic case'', the outflow should represent those seen in purely hydrodynamic simulations \citep[e.g,][]{Stone2009, Owen2014, Khodachenko2015, Tripathi2015, Carroll-Nellenback2016, McCann2019, Debrecht2020, Carolan2021a}.

In the magnetically controlled case, only a fraction of the planetary surface supports open field lines, and the mass loss rate is suppressed compared to the hydrodynamic case. 2D MHD simulations have suggested that surface magnetic field strengths as low as $\sim 0.3-3$ Gauss can suppress the mass loss rate of hot Jupiters by an order of magnitude \citep{Trammell2014, Owen2014, Khodachenko2015}. The critical magnetic field strengths required to suppress mass loss rates are smaller or comparable to many of the planetary magnetic field strengths in our Solar System (e.g. Jupiter's dipole moment is $\sim$4~Gauss, \citealt{Stevenson2003, Connerney2022}). Therefore it is important to consider the effects of magnetic fields on escape for exoplanets and understand how to determine their impact observationally. 3D MHD simulations that include the stellar wind also found the planetary mass loss rate decreases with planetary magnetic field strength \citep{Khodachenko2021}. However, we note similar simulations run by \citet{Carolan2021b} show the opposite dependence, and the differences are yet to be fully understood.

For lower-mass planets, strong planetary magnetic fields that suppress mass loss could significantly affect their evolution. \citet{Owen2019b} studied the effect that magnetic fields would have on the location of the radius valley. If these planets had surface magnetic field strengths $\gtrsim 3$ Gauss at the age of a few hundred Myrs\footnote{Since this is when the host stars are bright in the XUV, and thus when most of the mass-loss occurs, \citep[e.g.][]{Jackson2012}.}, then their cores would need to be ice-rich in order to match the observed radius valley. However, since the planets below the radius gap (that have well-constrained masses) generally have densities consistent with an Earth-like composition \citep{Dressing2015, Dorn2018,VanEylen2019}, they argue that this implies that the low-mass, close-in exoplanet population do not possess strong magnetic fields. Specifically, \citet{Owen2019b} argued that the dynamo-generated fields in the hydrogen/helium envelopes of mini-Neptunes must be weak ($\text{B} \lesssim 0.3$ Gauss) at young ages.

Unlike small planets, Hot Jupiters are thought to be stable to atmospheric mass loss independent of whether or not magnetic fields suppress escape \citep[e.g.,][]{Hubbard2007, Owen2013, Jin2014}. However, the strength of their magnetic fields is still of much interest. Hot Jupiters have radii larger than expected from interior structure models \citep{Laughlin2011,Thorngren2018}. In order to explain this discrepancy, these planets require an extra source of energy that must be deposited into the interior \citep[e.g,][]{Bodenheimer2001, Guillot2002, Arras2010, Youdin2010, Batygin2010,Komacek2017}. One popular explanation for this extra energy is Ohmic dissipation, in which the interaction of zonal winds and the planetary magnetic field generate electrical currents that resistively heat the interior \citep{Batygin2010, Perna2010, Batygin2011, Menou2012, Rauscher2013, Ginzburg2016}. Current modelling suggests that explaining the hot Jupiter radius anomaly in this way requires hot Jupiters to possess sizeable magnetic fields. 
For example, using 3D MHD simulations of the atmosphere of HD209458b, \citet{Rauscher2013} argue that the magnetic field must be greater than $3-10$ Gauss for Ohmic dissipation to be responsible for its inflated radius. {\tworc For a few hot Jupiters hosting stars, \citet{Cauley2019} found tentative evidence of magnetic star-planet interactions. From variations in the Ca II K line, they inferred these planets to have large magnetic fields $\sim$ 20-100 Gauss. However, we note that \citet{Cauley2019} caution the detected variability may be unrelated to the presence of a planet, and hence do not constrain planetary magnetic fields.} Independently, using a scaling law relating the magnetic energy density to the density and intrinsic luminosity of bodies ranging from Earth to low mass stars, \citet{Christensen2009} estimated the magnetic field of hot Jupiters to have magnetic field strengths of order 20-50~Gauss. In contrast, \citet{Griebmeier2004} argued that tidally locked hot Jupiters should have a much smaller magnetic field than Jupiter due to their slow rotation rate. 

To determine whether the effects of magnetic fields need to be accounted for in evolutionary modelling, we need to compare theoretical predictions of "magnetically controlled" and "hydrodynamic" outflows to direct observations of atmospheric escape. Additionally, these observations can be used to put constraints on the magnetic field strength of planets. The two most commonly used lines to directly observe atmospheric escape are Lyman-$\alpha$ \citep[e.g.,][]{Vidal-Madjar2003, LecavelierDesEtangs2010, Kulow2014} and He I 10830 \AA~ \citep[e.g.,][]{Spake2018, Allart2018, Nortmann2018}, which originates from the metastable $2^3S$ state of helium.     

\citet{Kislyakova2014, Ben-Jaffel2022} inferred the magnetic field properties of HD209458b and HAT-P-11b respectively, by matching the transit spectrum deduced from 3D outflow models to Lyman-$\alpha$ transits. {\rc In general, this approach is challenging because absorption in the interstellar medium generally obscures the line core \citep[e.g.][]{Landsman1993}, and therefore Lyman-$\alpha$ transits probe material at high radial velocities and distances far from the planet $\gtrsim 10 R_p$ where the outflow is affected by the circumstellar environment \citep[e.g][]{Owen2023a}. This causes observations to be degenerate with properties of the circumstellar environment, making it difficult to link observations to fundamental planetary quantities, such as the planetary magnetic field strength.}  

The helium line may provide a better tool in this endeavour. It probes material closer to the planet \citep[e.g,][]{Macleod2022} and observations can be done at high spectral resolution (R$\gtrsim$ 30,000) since it is accessible from the ground. This provides a unique opportunity to study the kinematics of the outflow near the planet. \citet{Oklopvcic2020} proposed using spectropolarimetry of this line to detect magnetic fields; however, this method is not currently feasible with 10m class telescopes but may be possible in the future with 30m class telescopes. 

Here, we propose a method for detecting the presence of magnetic fields strong enough to control the topology of the outflow using transmission spectroscopy in the 10830 \AA~ line of helium. {\rc Motivated by previous HD and MHD simulations, we suggest that "hydrodynamic" outflows (i.e. outflows where the planetary magnetic fields are dynamically unimportant) should lead to blue-shifted transmission spectra due to gas flowing from the hot day side to the cold night side (towards the observer during transit) as a result of pressure gradients from the anisotropic heating. In contrast, strong magnetic fields prevent the day-to-night outflow and cause the gas to move away from the observer during transit. As a result, we suggest that "magnetically controlled" outflows will produce transmission spectra that are (weakly) red-shifted.}   

\section{The Effect of Magnetic Fields on Photoevaporation}

In general, the planetary outflow is sensitive to both planet properties and the stellar environment. Following \cite{Owen2014}, one can coarsely partition the outflow into regimes based on the relative strength of the stellar magnetic field, stellar wind ram pressure, planetary magnetic field and planetary wind ram pressure. In this work, we consider regimes where either the ram pressure of the planetary wind or the planetary magnetic field pressure dominates the stellar components close to the planetary surface, where most of the helium absorption occurs. We can estimate whether the outflow is magnetically controlled by considering the dimensionless quantity $\Lambda$, which is the ratio of the ram pressure of the wind to the magnetic pressure of the planet. For a planet with a dipolar magnetic field, this ratio is equal to     

\begin{align}
    \Lambda_{\text{p}} = \frac{2\dot{M}vr^4}{B_0^2R_P^6}
\end{align}

where $\dot{M}$ is the outflow rate, $B_0$ is the surface magnetic field strength of the body, $R_P$ is the radius of the body and $r$ is the distance to the centre of the body. When $\Lambda_{\text{p}} \ll 1$, the flow is magnetically controlled, and the outflowing gas is forced to follow magnetic field lines. When $\Lambda_{\text{p}} \gg 1$, the outflow will drag magnetic field lines and should follow a flow structure similar to a pure hydrodynamic outflow. Using an ``energy-limited'' outflow \citep[e.g.][]{Baraffe2004} where

\begin{align}
\label{eq:el_mlr}
\dot{M} = \eta\frac{\pi F_{\text{XUV}}R_P^3}{4GM}
\end{align}

\noindent we estimate the threshold magnetic field such that magnetic pressure and ram pressure are equal (i.e $\Lambda_p = 1)$. For fiducial parameters appropriate for a typical hot Jupiters, this is: 

\begin{align}
\begin{split}
    B \approx &~ 0.015 \left(\frac{\eta}{0.1}\right)^\frac{1}{2}\left(\frac{F_{\text{XUV}}}{10^4\text{ erg s}^{-1}}\right)^\frac{1}{2}\left(\frac{v}{10\text{ km s}^{-1}}\right)^\frac{1}{2}\left(\frac{M_p}{M_J}\right)^{-\frac{1}{2}}\\
    &\times \left(\frac{R_p}{R_J}\right)^\frac{1}{2}\left(\frac{r}{R_p}\right)^2 \text{ Gauss}
\end{split}
\end{align}

\subsection{Post-processing existing simulations} \label{Geometry}

To assess whether "hydrodynamic" and "magnetically controlled" outflows will produce blue- and red-shifted transit spectra, respectively, we want to compare the synthetic transit spectra produced from simulations. There are no 3D radiation-hydrodynamics photoevaporation simulations that include helium and magnetic fields. Given the complexity of post-processing simulations (as detailed in Section~\ref{sec:adding_he}), which would be difficult in 3D, as a first step, we choose to use existing 2D photoevaporation simulations of a hot Jupiter presented in \citet{Owen2014}. These radiation-hydrodynamic simulations did and did not include magnetic fields for the same setup and are ideal for our experiment. To analyze the case where the ram pressure of the planetary outflow dominates over the magnetic field of the planet ($\Lambda_{\text{p}} > 1)$, we use a purely hydrodynamic simulation of the outflow. In the opposite case, where magnetic pressure dominates over ram pressure of the planetary outflow ($\Lambda_{\text{p}} < 1$), we use MHD simulations of the outflow.   

In these simulations, the planet has a dipolar magnetic field. The planetary dipole was chosen to be parallel to the orbital angular momentum vector of the planet. There are currently no constraints on the geometry of the magnetic field of exoplanets. However, since these planets are very close to their host stars, they are often assumed to be tidally locked. Thus, as a starting point, it is sensible to guess that any dynamo-generated magnetic field may be similar to Jupiter's or Saturn's and (almost) points along the rotation axis.

In some of the simulations, an additional component corresponding to the stellar magnetic field was added. This contribution is modelled as a constant field that points along the pole of the planet:   

\begin{align}
B_{*} = \beta_{*}B_{p}\hat{z}  
\end{align}

where $\beta_{*}$ controls the relative strength of the stellar and planetary magnetic fields. 

The original simulations were performed on a spherical polar grid $(r, \theta)$, which we can convert to Cartesian coordinates $(x, z)$, where the $x$ axis points along the line joining the star and the planet and the $z$ axis is parallel to the planet's orbital angular momentum axis. It is important to state that atmospheric escape is generally a 3D process, and therefore, several assumptions need to be made in order to be able to perform sensible 2D simulations. Firstly, the planet needs to be tidally locked to have a constant day and night side. At the same time, rotational effects are ignored. Inside the Hill sphere, these effects are small \citep{Murray-Clay2009} and can be ignored; however, they are important for shaping the flow on the largest scales \citep[e.g,][]{McCann2019}. Since absorption by helium generally comes from close to the planet, this approximation is reasonable.     

In the hydrodynamic case, these approximations mean that the outflow is rotationally symmetric around the line joining the planet to the sub-stellar point. The presence of this magnetic field stops this setup from being rotationally invariant, such that the flow is no longer globally axisymmetric. Therefore, the simulations required each cell to locally have the property that $\partial_{\phi} = 0$. Since the magnetic field suppresses the azimuthal flow, this approximation is realistic. Thus, the simulation can be considered a thin slice of the domain in the x-z plane. Since the magnetic field predominately governs the outflow geometry, we expect the flow structure on the day side to be similar as we rotate around the dipole axis. On the night side, the outflow is suppressed since heat is not efficiently transported across magnetic field lines. As a result, only the day side of the atmosphere was simulated (see sec 5.1 of \citealt{Owen2014} or \citealt{Trammell2014})  

\subsection{Addition of Helium} \label{sec:adding_he}
    
The lifetime of the metastable triplet level in the outflow is determined by the rates of collisional, radiative and photoionisation processes that depopulate it. As radiative transitions into the ground state are highly suppressed \citep{Drake1971}, depopulation of this state is dominated by transitions due to collisions and photoionization \citep[e.g.,][]{Oklopvcic2018, Oklopvcic2019}. In the outflow, the lifetime of the metastable state may be comparable to the flow timescale (see Figure 4 of \citealt{Oklopvcic2019}). Thus, the helium metastable level is often \emph{not} in local statistical equilibrium. This means traditional post-processing techniques of simply using the gas' local properties to compute absorption cross-sections are inappropriate. Therefore, we must compute the level populations by solving the time-dependent radiative transfer equations for gas parcels moving in the outflow. 

We assume that the gas is composed of solar abundances of hydrogen and helium; however, we note that the mass fraction of helium in the outflow can be adjusted depending on how efficiently helium is expected to be dragged in the predominantly hydrogen flow \citep[e.g.][]{Malsky2020}. Helium is not expected to contribute to the cooling of the gas, and therefore the predominant way that it affects the outflow is by raising its mean molecular weight. For a solar abundance of helium, the sound speed of the gas can be modified $\sim$ 25\%. Ultimately, though, adding helium should not have a large effect on the geometry of the outflow. Therefore, it is reasonable to consider the pure flow structure of the pure hydrogen outflow simulated by \cite{Owen2014} as representative of a hydrogen-helium outflow. Similarly, the distribution of helium ions and atoms in each electronic state has little impact on the outflow structure.

\section{Methods}

\subsection{Helium Distribution in the Outflow}
\label{sec:maths} 

The first task is to calculate the distribution of helium atoms in the outflow. We track the fraction of helium in the neutral singlet and triplet states along streamlines in the outflow. We only track the number of neutral helium atoms in the ground state $1^1S$ and metastable $2^3S$ state since excited states quickly radiatively decay to one of these two states with a matching singlet/triplet configuration. Our problem has two parts: (i) determining streamlines in the flow and (ii) calculating the fraction of helium in the triplet state along these streamlines. 

In the case of the pure hydrodynamic flow, we solve for the 2D position of a streamline as a function of length, $l$, down the streamline using equations:    

\begin{align}
    \frac{dr}{dl} &= \frac{u_r}{u} \label{eq:streamline_hydro_r}\\
    \frac{d\theta}{dl} &= \frac{u_{\theta}}{ru} 
    \label{eq:streamline_hydro_th}
\end{align}

We then derive the steady state distribution of helium atoms along streamlines using Equations \ref{eq:f1_eq} and \ref{eq:f3_eq}, which account for the major transitions in and out of the singlet and triplet states of helium \citep[][]{Oklopvcic2018}.   

\begin{align}
\label{eq:f1_eq}
\begin{split}
u\frac{df_1}{dl} = \, &(1 - f_1 - f_3)n_e\alpha_1 + f_3A_{31} + f_3q_{31a}n_e\\
&+ f_3q_{31b}n_e + f_3Q_{31}n_{HI} - f_1\phi_1 e^{-{\tau_1}} - f_1q_{13}n_e 
\end{split}
\end{align}

\begin{align}
\label{eq:f3_eq}
\begin{split}
u\frac{df_3}{dl} = \, &(1 - f_1 - f_3)n_e\alpha_3 - f_3A_{31} - f_3q_{31a}n_e\\
&- f_3q_{31b}n_e - f_3Q_{31}n_{HI} - f_3\phi_3 e^{-{\tau_3}} + f_1q_{13}n_e 
\end{split}
\end{align}

\noindent $f_1$ and $f_3$ are neutral helium's singlet and triplet fractions, respectively. $n_e$ and $n_{HI}$ are the number density of electrons and hydrogen atoms. $\phi_1$ and $\phi_3$ are the photoionization rate for the singlet and triplet states, respectively. We calculate the photoionisation rate for the singlet and triplet states by assuming that all the photoionizing flux is concentrated at a photon energy $\overline{h\nu}$ of $24.6\text{ eV}$ and $4.8 \text{ eV}$ respectively.   

\begin{align}
    \phi_{1} = \int_{24.6\text{eV}}^{\infty}\frac{F_{\nu}}{h\nu}\sigma_{1}(\nu)d\nu \approx \frac{F_{\overline{\nu}}}{\overline{h\nu}}\sigma_{1}(\overline{\nu})
\end{align}

\begin{align}
    \phi_{3} = \int_{4.8 \text{eV}}^{13.6\text{eV}}\frac{F_{\nu}}{h\nu}\sigma_{3}(\nu)d\nu \approx \frac{F_{\overline{\nu}}}{\overline{h\nu}}\sigma_{3}(\overline{\nu})
\end{align}

Based on the spectral energy distribution for late K stars from the MUSCLES survey \citep[]{MUSCLES1, Loyd2016, Youngblood2016}, we estimate the total singlet ionising flux to be 0.2 of the hydrogen ionizing flux and the total triplet ionising flux to be about the same hydrogen ionizing flux. As the simulations were run at high UV fluxes $\sim 10^6 \text{ erg cm}^{-2}\text{ s}^{-1}$, we set $F_{\overline{\nu}} = 2 \times 10^5 $ and $ 10^6 \text{ erg cm}^{-2}\text{ s}^{-1}$ for the singlet and triplet states respectively. We varied these fluxes by an order of magnitude in both directions and found that it did not qualitatively affect our results. $\sigma_{i}$ is the photoionization cross-section of the state. The photoionisation cross section for these photons are $5.48 \times 10^{-18} \text{ cm}^2$ \citep{Hummer1998} and $7.82 \times 10^{-18} \text{ cm}^2$ \citep{Brown1971} respectively. In this work, we set both $\tau_1 = 0$ and $\tau_3 = 0$. Since hydrogen also absorbs $24.4\text{ eV}$ radiation, we expect the $\tau = 1$ surface to $24.4 \text{eV}$ photons to be above the $\tau = 1$ surface to $4.8 \text{eV}$ photons (and close to the hydrogen ionization front in the simulation). Since the main pathway to populate the triplet state is by photoionization of the singlet state and then recombination into the triplet state, we expect that this will underestimate the triplet fraction below the $\tau = 1$ surface to $24.4 \text{eV}$. However, since the hydrogen ionization front is deep in the atmosphere of the planet (in these simulations, it is $\sim 1.03 \text{ R}_{p}$), the gas below this surface contributes minimally to the overall transit signal. 

The recombination and collisional excitation/de-excitation coefficients depend on the temperature of the gas. As the simulations were run for high UV fluxes $(\gtrsim 10^6 \text{ erg cm}^{-2}\text{ s}^{-1})$, the gas is in the radiative-recombination regime \citep{Murray-Clay2009} and therefore thermostats to $\sim 10^4 \text{ K} $ above the ionization front. As a result, we use all coefficients at $10^4 \text{ K}$. $\alpha_1 = 2.16 \times 10^{-13} \text{ cm}^{3}\text{ s}^{-1}$ \& $\alpha_3 = 2.25 \times 10^{-13} \text{ cm}^{3}\text{ s}^{-1}$ are the case A recombination coefficients for the singlet and triplet state respectively \citep{Osterbrock2006}. The electron density ($n_e$) is calculated in the \citet{Owen2014} simulations,  which only accounted for electrons produced by the ionisation of hydrogen atoms since this is the dominant source. $A_{31} = 1.272 \times 10^{-4} \text{ s}^{-1}$ is the radiative transition rate between the metastable triplet state and the ground singlet state \citep{Drake1971}. The transitions between the triplet and singlet states due to collisions with electrons are denoted by the temperature dependant coefficients: 

\begin{align}
q_{ij} = 2.1 \times 10^{-8} \sqrt{\frac{13.6 \text{ eV}}{kT}}\text{exp}\left(-\frac{E_{ij}}{kT}\right)\frac{\Upsilon_{ij}}{w_i} \text{ cm}^{3}\text{s}^{-1}
\end{align}
where $E_{ij}$ is the energy difference between the states from \citet{Martin1973}, $w_i$ is the statistical weight of the state and $\Upsilon_{ij}$ is the effective collision strength from \citet{Bray2000}. At $10^4$ K, the two transitions from the metastable state to singlet levels are given by $q_{31a} = 2.7 \times 10^{-8} \text{ cm}^3 \text{ s}^{-1}$ and $q_{31b} = 5.2 \times 10^{-9} \text{ cm}^3 \text{ s}^{-1}$. The transition from the ground singlet state to the metastable state is given by $q_{13} = 5.7 \times 10^{-19} \text{ cm}^3\text{ s}^{-1}$. Finally, transitions out of the metastable state due to collisions with hydrogen atoms via associative ionisation and Penning ionisation have a combined rate of $Q_{31} \sim 5 \times 10^{-10} \text{ cm}^3 \text{ s}^{-1}$ \citep{Roberge1982}. 

\begin{figure*}
\centering
\includegraphics[]{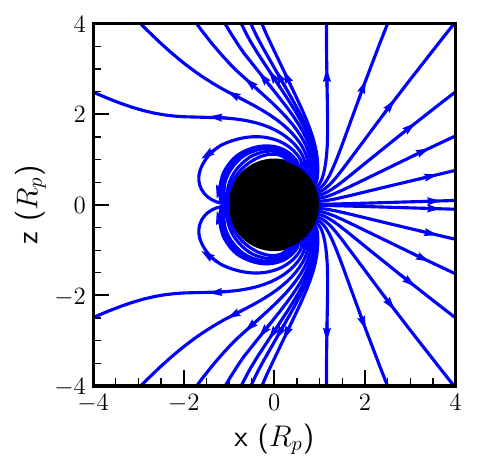}
\includegraphics[]{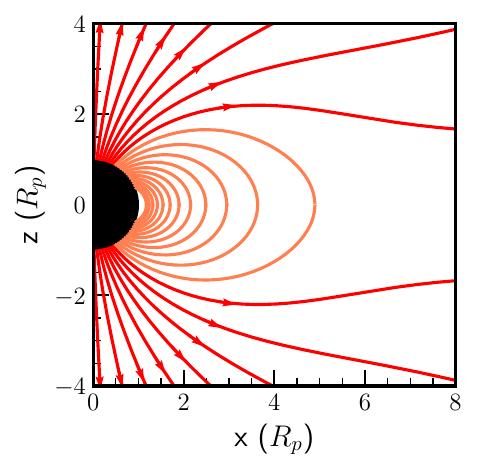}
\caption{Left: Streamlines of the outflow for the case of a planet without a magnetic field. The star is located to the right of the planet along the line z = 0. The planet is shown by a black circle centred on the origin. The orientation is the same for all subsequent 2D plots. {\rc Due to anisotropic heating, the day side is much hotter than the night side. The resulting pressure gradients cause the outflow to wrap around the planet, creating gas travelling towards an observer in transit in the transmission region.} Right: Magnetic field lines for a planet with a surface magnetic field strength of 3 Gauss. Orange lines are closed field lines where the magnetic field is strong enough to confine the flow. The red lines are open magnetic field lines, which also are gas streamlines. The direction of the arrows shows the direction of the outflow. In this case, the magnetic field prevents the day-to-night side outflow and causes the gas to move away from the observer in transit in the transmission region.}
\label{fig:streamlines}
\end{figure*}

Equations \ref{eq:streamline_hydro_r}-\ref{eq:f3_eq} form a set of coupled first-order differential equations. The streamlines are initialised at a constant radius $1.02 \text{ R}_{p}$ with an initial singlet and triplet fraction that is in equilibrium at 500 K with $\phi_1 = \phi_3 = 0$. We solve these differential equations numerically using the implementation of the LSODA method \citep{Petzold1983} provided in the \textit{scipy} library \citep{scipy}. For numerical stability, we solve \ref{eq:f1_eq} and \ref{eq:f3_eq} as function of log($l$). The absolute and relative tolerances used are $1 \times 10^{-13}$ and $1 \times 10^{-10}$, respectively and the maximum step size chosen is $1 \times 10^{-3}$. All gas properties required to solve these equations (e.g density, velocity etc.) were interpolated from the original simulations using bivariate spline interpolation over the polar grid. To calculate the metastable helium fraction at any point in the outflow, {\rc we interpolate the values for fifty streamlines, linearly spaced in angle, originating on the planet's day side. Testing indicates this is sufficient to calculate the transmission spectrum accurately.}  Figure \ref{fig:streamlines} shows how streamlines originating on the planet's day side wrap around to the night side.

In the MHD case, we have to consider both the field centred on the poles where the magnetic field lines are opened by the outflow and a region centred on the equator where magnetic pressure dominates over thermal pressure so that the field lines are closed. Since the fluid is required to follow magnetic field lines (and magnetic field lines are dragged along by the fluid), we can determine the streamlines of our gas in the open region by tracing magnetic field lines. This is helpful because we can simultaneously determine the region in which the field lines are closed. These are calculated analogously to \ref{eq:streamline_hydro_r} and \ref{eq:streamline_hydro_th} replacing velocity with magnetic field:    

\begin{align}
    \frac{dr}{dl} &= \frac{B_r}{B}\\
    \frac{d\theta}{dl} &= \frac{B_{\theta}}{rB} 
\end{align}
where $B_r$ and $B_{\theta}$ are the radial and angular components of the magnetic field, respectively. Figure \ref{fig:streamlines} shows the magnetic field lines for a fiducial planetary magnetic field strength of 3 Gauss. We use this value of the planetary magnetic field to illustrate the magnetically controlled case throughout the rest of the paper.         

In the regions where the streamlines are open, we solve the steady state distribution of helium atoms using Equations \ref{eq:f1_eq} and \ref{eq:f3_eq} as in the hydrodynamic case. In the regions in which they are closed, the gas is approximately in magneto-hydrostatic equilibrium, and as a result, the helium-level populations are in statistical equilibrium. Therefore we solve for the distribution of helium atoms using Equations \ref{eq:f1_eq} and \ref{eq:f3_eq} by setting the advection term (the left-hand side) to zero. The method used to solve these equations numerically is the same as the hydrodynamic one, with the addition of a step to check whether the field line is open or closed to decide to solve the non-equilibrium or equilibrium equation for the distribution of helium atoms.

\begin{figure*}
\centering
\includegraphics[width = 0.45\textwidth]{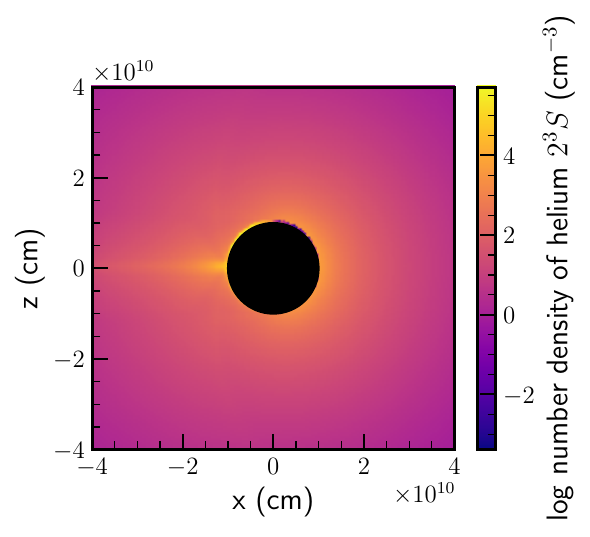}
\includegraphics[width = 0.45\textwidth]{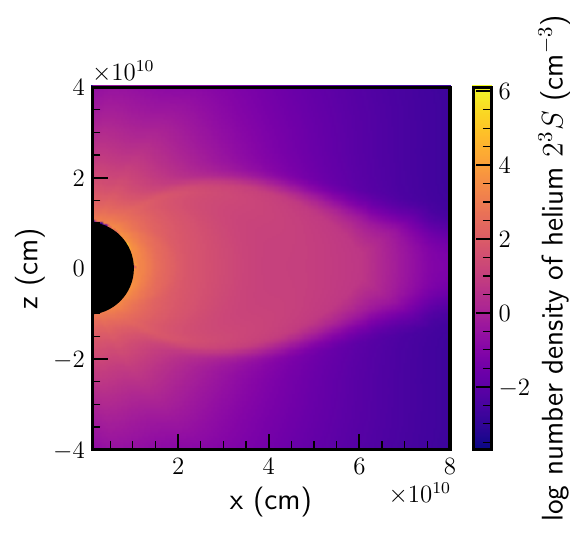}
\caption{The number density of helium in the $2^3S$ state in the outflow for the hydrodynamic case (left) and the magnetically controlled case (right). In the hydrodynamic case, the density appears to be fairly spherically symmetric. In the magnetically controlled case, the region with closed magnetic field lines has a higher number density of metastable helium atoms than the outflowing regions. This is primarily because the total number density of gas in this region is higher, as it is in hydrostatic equilibrium.}
\label{fig:density_mhd}
\end{figure*}

In Figure \ref{fig:density_mhd}, we show the number density profiles of helium in the triplet state for the hydrodynamic and magnetically controlled outflow.   

\subsection{Computing Synthetic Transit Profiles}

Our goal is to calculate the absorption of stellar radiation at $10830$ \AA~ due to the presence of the escaping helium. To elucidate the basic principles, we produce synthetic transmission spectra at mid-transit for the planet transiting with an impact parameter of zero. Since the simulations used to generate the densities and velocities of the outflow are 2D, it's only at mid-transit and with an impact parameter of zero that our approximate symmetry is applicable. 

With the assumptions described in \ref{Geometry}, the outflow in the purely hydrodynamic case is rotationally invariant around the line connecting the star and planet (x-axis). This symmetry means that we can identify the z coordinate of our grid with the radial coordinate on the stellar disc. Therefore, the optical depth of the outflow for a ray emanating at any point on the stellar disc is only a function of z and is given by Equation \ref{eq:tau_c_eq}. This can be integrated over the stellar disc to give the excess absorption caused by the escaping gas.   

\begin{align}
\tau_{\nu}(z) = \int_{-\infty}^{\infty}n_3(x, z)\sigma_{\nu}(u_x, T)dx
\label{eq:tau_c_eq}
\end{align}

{\tworc
\begin{align}
\text{Excess Depth}(\nu) = \frac{2\int_{R_p}^{R_s}e^{-\tau_{\nu}(z)}zdz + R_{p}^2}{R_s^2}
\label{eq:F_c_eq}
\end{align}}
 
\noindent where $n_3$ is the number density helium atoms in the metastable state, $R_p, R_s$ are the radius of the planet and star, respectively and $\sigma_{\nu}$ is the absorption cross-section of the helium to photons at a frequency $\nu$, which is a function of the line of sight velocity of the gas, $u_x$, and the temperature. The absorption cross-section is given by 

\begin{align}
    \sigma_{\nu} = \frac{\pi e^2}{m_{e}c}f\Phi(u_x, T)
\end{align}

where $e$ and $m_e$ is the charge and mass of the electron, $c$ is the speed of light, $f$ is the oscillator strength of the transition and $\Phi$ is the Voigt line profile. The Gaussian part of the Voigt profile has a standard deviation of $\sigma$; the Lorentzian part has a half-width at half-maximum of $\gamma$, and the line centre has been Doppler shifted to $\nu$.  

\begin{align}
    \sigma = \sqrt{\frac{k_bT}{m_{He} c^2}}\nu_0 ,\quad \gamma = \frac{A}{4\pi}, \quad \nu = \nu_0\left(1 - \frac{u_x}{c}\right)
\end{align}

Here $\nu_0$ is the rest frequency of the helium line, and $A = 1.022 \times 10^7 \text{ s}^{-1}$ is the Einstein $A$ coefficient taken from the NIST atomic database \citep{NIST_ASD}. It is worth noting that the $2^3S \rightarrow 2^3P$ transition in Helium actually consists of three distinct lines at 10830.34 \AA, 10830.25 \AA~ and 10829.09 \AA~due to fine structure splitting. The first two of these transitions are blended; however, the third can be distinguished at the spectral resolution of current instruments. The oscillator strength for these transitions, taken from the NIST atomic database \citep{NIST_ASD}, are given by 0.300, 0.180, and 0.060, respectively. The absorption cross-section at any wavelength is calculated by summing up the contribution from these three transitions. 

Similar to \citet{Oklopvcic2018}, when computing the optical depth, we only took into account gas within the Hill sphere of the planet, which we choose to have a radius of $4 R_p$. This approximately corresponds to the Hill sphere radius of a Jupiter-mass planet around a solar-mass star at 0.04 AU. Beyond the Hill sphere, the outflow is significantly affected by the Coriolis force, the stellar tidal field and the stellar wind. Here, the model is no longer applicable. However, unless the stellar wind is very strong, these regions do not contribute greatly at mid-transit \citep[e.g.,][]{Macleod2022}. If the planet is further away from the star or the magnetic field is strong, the planetary outflow may not be affected by the circumstellar environment at larger distances from the planet. In Appendix \ref{appendixA}, we provide transit spectra taking into account gas up to $8 R_p$ to illustrate this effect.  

To do this calculation, we construct a linearly spaced $300 \times 600$ cell grid in the $(x, z)$ plane, with x ranging from [$-4R_p$, $4R_p$] and z ranging from [$R_p$, $4R_p$]. The integrals in \ref{eq:tau_c_eq} and \ref{eq:F_c_eq} are computed with this grid using a middle Riemann sum.        

The method we use in the magnetic case is different. As discussed above, the rotational symmetry around the line connecting the star and planet (x-axis) present in the purely hydrodynamic case is broken by the dipolar structure of the magnetic field. Therefore, we cannot simply compute the 3D density structure by rotating the 2D density distribution around the x-axis. The presence of a planetary magnetic field shuts off the outflow on the night side planet since heat cannot be transported effectively across magnetic field lines. Therefore, we set the density of helium on the night side of the planet to zero. On the day side, we build the 3D density distribution by rotating the 2D density of helium distribution around the z-axis (symmetry axis of the dipole). {\rc In order to account for the reduction of EUV flux incident on the atmosphere away from the substellar line due to geometrical effects, we scale the number density of helium in the triplet state in each 2D slice by the square root of the reduction of incident EUV flux at the base the slice. We do this because the simulations are done at high EUV fluxes such that the outflow is in ionization-recombination equilibrium. In this regime, the temperature of the outflow thermostats to $10^4$ K and the density at the base of the outflow is proportional to the square root of the incident EUV flux \citep[e.g][]{Murray-Clay2009,Owen2016}.} 

To perform synthetic observations, we create a 3D cylindrical grid from the Cartesian product of a 2D grid on the stellar disc and a 1D grid along the line connecting the star and planet. The 1D grid has 150 cells and ranges from [$0$, $4 \text{R}_{p}$]. The 2D grid on the stellar disc has 400 cells. We calculate the optical depth of the outflow at the centre of each cell on the stellar disc by integrating through the 3D density distribution using a middle Riemann sum. This can then be translated into an excess depth by integrating over the stellar disc using a double Riemann sum:   
{\tworc
\begin{align}
    \text{Excess Depth}(\nu) = \frac{\sum_{j}e^{-\tau_{\nu,j}}A_j + \pi R_p^2}{\pi R_s^2} 
    \label{excess_depth_mhd}
\end{align}
}

\noindent where $A_j$ is the area of the $j^{th}$ cell.

\begin{figure*}
\centering
\includegraphics[width = 0.45\textwidth]{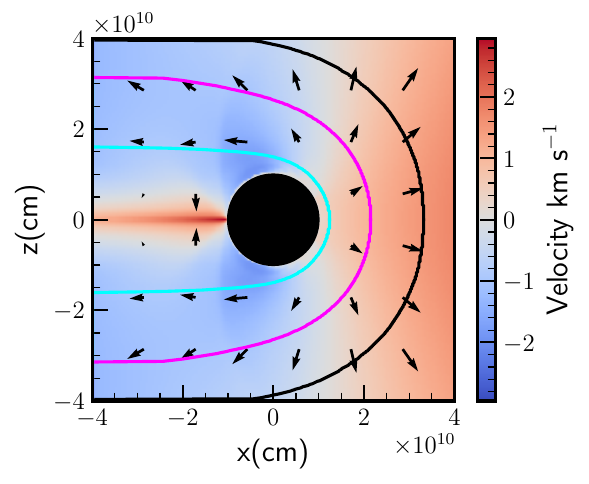}
\includegraphics[width = 0.45\textwidth]{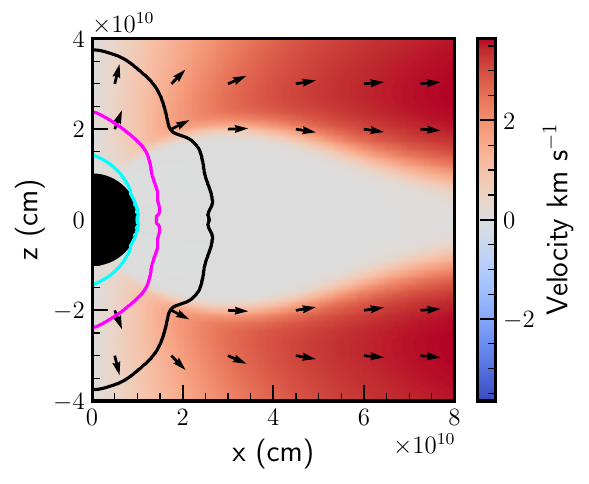}
\caption{Colour map of the line of sight velocity of the outflowing gas in the hydrodynamic case (left) and the magnetically controlled case (right). Contours mark the line of sight optical depth integrated into a two-angstrom bin around the peak absorption wavelength. The black, magenta and cyan mark the $\tau = 0.01, 0.1 \text{ and } 1$ contours, respectively. The arrows show the velocity field.}
\label{fig:tau_contours_hd}
\end{figure*}

\section{Results}

\begin{figure}
\centering
\includegraphics[width = 0.45\textwidth]{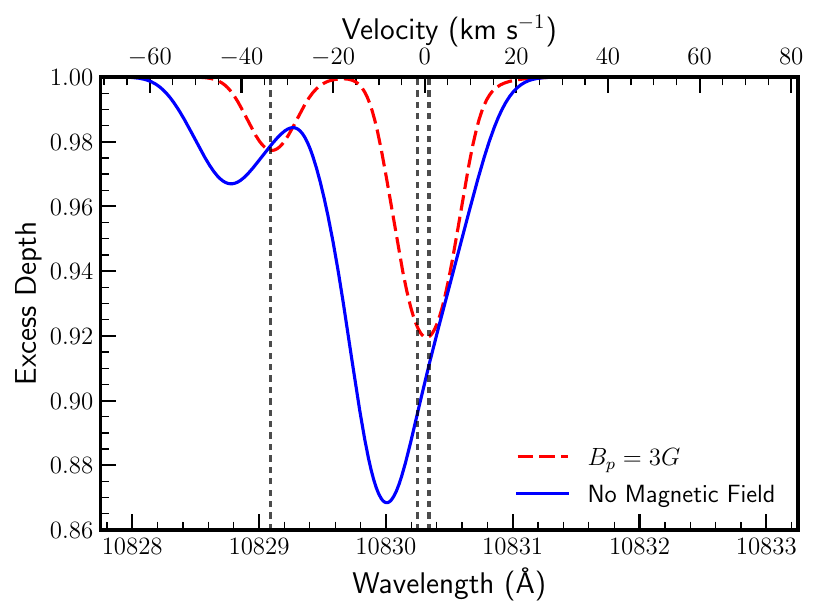}
\caption{Calculated excess absorption for a planet with i) no magnetic field (solid blue) and ii) a surface magnetic field of $3$ Gauss (dashed red) at mid-transit with an impact parameter of 0. The velocity scale is centred on the blended transitions at 10830.25 and 10830.34 \AA}
\label{fig:hdvsmhd}
\end{figure}

Figure \ref{fig:hdvsmhd} shows the excess transit depth for the hydrodynamic and magnetically controlled outflows. As expected, we find the hydrodynamic outflow leads to a blue-shifted spectrum. Figure \ref{fig:tau_contours_hd} shows that most of the absorption comes from the terminator, where gas is moving towards the observer in transit. The peak absorption is at $-8.3 \ \text{km s}^{-1}$, which is of order the sound speed of the outflowing gas. It is worth noting that we still see significant absorption in the red wing due to absorption from gas moving towards the star (away from the observer in transit), which has originated from low latitudes on the planet. As a result, the spectrum is asymmetric and broader than expected from thermal broadening.      

In contrast, a planet with a $3$ Gauss magnetic field shows a symmetric transit spectrum, with a peak absorption centred at $0.5$ km $\text{s}^{-1}$. Most of the absorption comes from the dead zone, where the velocity of the gas is stationary, and an area near the poles, where the line of the sight velocity is small (see Figure \ref{fig:tau_contours_hd}). As a result, the spectrum only shows a small velocity shift. Furthermore, since almost all the absorption comes from gas at low velocities, the width of the spectrum is primarily set by the thermal velocity of the gas. 

\begin{figure}
\centering
\includegraphics[width = 0.45\textwidth]{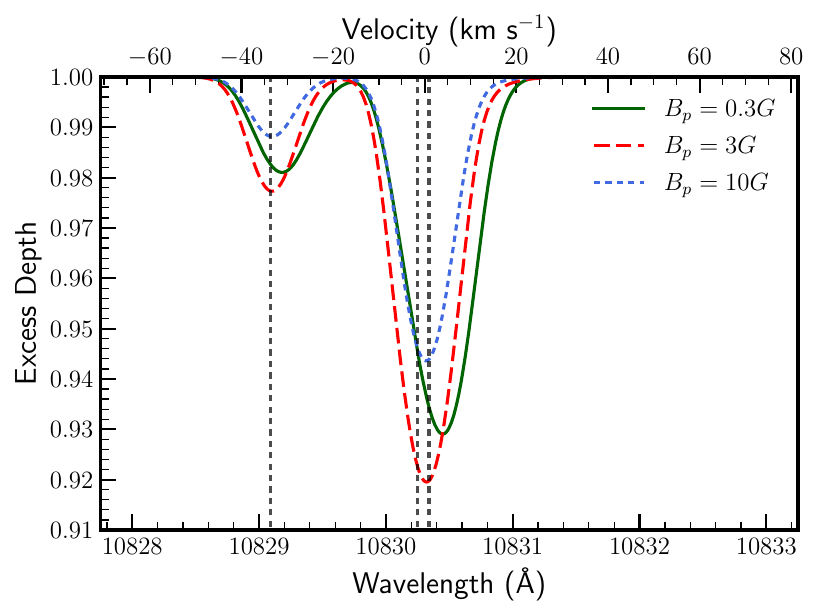}
\caption{Calculated excess absorption as a function of wavelength for different planetary magnetic field strengths. For a planetary magnetic field of 3 Gauss, we also include a case in which the stellar magnetic field is non-zero. The velocity scale is the same as in Figure \ref{fig:hdvsmhd}.}
\label{fig:mhd_transits}
\end{figure}

In Figure \ref{fig:mhd_transits}, we show how the transit spectrum varies with planetary magnetic field strength. Interestingly, the transit depth does not decrease monotonically with magnetic field strength, even though the fraction of open field lines (and the mass loss rate) decreases smoothly as a function of magnetic field strength. As the magnetic field strength is increased, the size of the region where the magnetic field lines are closed increases. This region then blocks more of the stellar disc, contributing more to the transit depth. At the same time, since the mass loss decreases, the column density of gas in the outflowing regions near the poles decreases, contributing less to the transit depth. The weight of these contributions has an opposite dependence on the strength of the magnetic field and results in the transit depth not being a monotonic function of the magnetic field strength. 
 
The overall velocity shift of the transmission spectrum also varies as the magnetic field strength changes. At lower magnetic field strengths, gas pressure is able to open up field lines at lower latitudes. The gas coupled to these field lines has a higher line sight velocity, resulting in a red-shifted spectrum. In the $0.3$ Gauss case, shown in Figure \ref{fig:mhd_transits}, this velocity shift is $3.8 \text{ km s}^{-1}$. 

For the $3$ Gauss case, we also include a situation where the stellar magnetic field is non-zero, and the ratio of stellar to planetary magnetic field strength is given by $\beta_{*} = 0.001$. For a planet at 0.04 AU around a solar radius star, this corresponds to a stellar surface magnetic field of 1 Gauss. In general, the stellar magnetic field is able to open out more field lines. As mentioned above, this can either increase or decrease the transit depth depending on how the contributions from the closed region and outflowing regions change. However, in this case, we actually see no change in the transit spectrum. This is because the opening of field lines is completely dominated by the outflow rather than the stellar field. This effect is explained in Figure 4 of \citet{Owen2014}. 

\section{Discussion and Conclusions}

We have post-processed 2D HD and MHD simulations of a hot Jupiter undergoing photoevaporation and calculated in-transit absorption signal due to the 10830 \AA~ line of helium. If the planet has no magnetic field, the transit signal is blue-shifted $\Delta v = -8.3 \text{ km s}^{-1}$  due to the absorption from gas that flows from the hot day side to a cold night side. When a planetary magnetic field (between 0.3-10 Gauss) is included, the transit signal has either no apparent velocity shift or a very slight red-shift, as escaping gas from the day side is tied to magnetic field lines and moves away from the observer in transit.      

At the current resolution of ground-based instruments, this difference in line-centre velocity is observable and may be used to test whether outflows are magnetically controlled or not and consequently put constraints on the magnetic field strengths of planets. As the observed velocity difference is due to large-scale changes in the geometry of the flow, this method should be fairly robust to the specifics of the photoevaporative flow (e.g. mass loss rate, temperature). However, further work is required to ensure that our results are still true in full 3D simulations over a more extensive region of parameter space and for various magnetic field configurations. We will discuss this in section \ref{3D Effects}.      

For magnetically controlled flows, observationally distinguishing between different planetary magnetic field strengths is a more difficult task. The clearest difference is that lower field strengths with more open field lines lead to a more red-shifted transit. More open field lines can also increase high-velocity gas absorption far from the planet, leading to asymmetric transits with an extended red wing. This is only visibly apparent in the 0.3 Gauss case when we consider gas within 8$R_p$ instead of $4R_p$ (see Figure \ref{fig:mhd_transits8Rp} in Appendix A). Observations of the escaping atmosphere of WASP-52b show that the transit spectrum is unshifted with respect to the planet's rest frame and has a slight asymmetry in the red wing. We speculate that this might indicate that the outflow is magnetically controlled. Overall, we caution that since the differences in these spectra are relatively minor, model and observational uncertainties most likely make different magnetic field cases indistinguishable at the resolution of current instruments.

\subsection{Limitations}

\subsubsection{Use of \citet{Owen2014} simulations and 3D effects} \label{3D Effects}

In reality, atmospheric escape is a 3D dimensional process, and 3D radiation magneto-hydrodynamics simulations that include rotational effects are necessary to confirm the validity of this work. The \citet{Owen2014} simulations are simplistic when compared to current state-of-the-art simulations; however, as discussed above, they are the only simulation set that included radiative transfer, models with and without magnetic fields and a reasonable parameter study within the same framework. This allowed us to make meaningful comparisons between simulations that did and did not include magnetic fields. However, it is important to discuss some of the limitations of these simulations so that our results are non-conflated with specific quantitative results arising from our post-processing of this simulation set. The \citet{Owen2014} simulations were the first multi-dimensional simulations of photoevaporation that included on-the-fly radiative transfer.  As such, they used a simplified algorithm which assumed an ionization-recombination balance for the hydrogen and took the ionized gas to be at $10^4$~K, where the temperature was computed based on the local ionization fraction \citep[e.g.][]{Gritschneder2009}. This ionization-recombination balance is well known to be applicable to planets around young stars \citep[e.g][]{Murray-Clay2009,Owen2016}, yet most observations of helium outflows are for old planets where this approximation does not hold. The approximations of the \citet{Owen2014} simulations are likely driving two quantitative results of our simulations: firstly, the high fluxes produce a high rate of recombinations, and thus, our helium transit depths are likely overestimated compared to older systems; secondly, the outflow velocity (and hence the magnitude of the velocity shift) depends directly of the gas' temperature, with higher temperatures giving higher outflow velocities. Outflows around lower-mass, older planets are more likely to be in the ``energy-limited'' regime \citep{Owen2016}, where the outflow temperatures are necessarily lower, and hence we suspect our simulated $\sim 8$~km~s$^{-1}$ blue-shifts are likely overestimates. However, as discussed above, based on the results of more recent 3D simulations, we are confident our qualitative conclusions about blue-shifts arising in the hydrodynamic cases are robust. Hence, the detection or lack of blue shift can be used to make inferences about the strength of any planetary magnetic field. 

We also re-emphasize that we focus on a particular area of parameter space where the ram pressure of the planetary wind or the magnetic pressure of the planet dominates over the stellar magnetic pressure or stellar wind ram pressure close to the planet. Whether the stellar wind or stellar magnetic field has a larger influence on the orbit of the planet depends on the mass loss rate of the star, the magnetic field of the star, stellar wind velocity and the semi-major axis of the planet. For hot Jupiters, both the case where the stellar magnetic field or the case where stellar wind dominates is possible for sensible stellar and planetary parameters. As the magnetic pressure for a dipolar field drops $\sim r^{-6}$ compared to the stellar wind ram pressure $r^{-2}$, the stellar wind is more likely to dominate for sub-Neptunes which are generally further from their star than hot Jupiters.

The primary effect of increasing the strength of the stellar magnetic field is to open more planetary field lines. At low stellar magnetic field strengths, this can increase the mass loss rate. At high stellar magnetic field strengths, the flow cannot smoothly transition through the sonic point along these open field lines, and the mass loss rate can actually decrease \citep{Owen2014}. These changes can vary the transit depth. However, we don't expect it to significantly change the velocity of the absorbing gas and, therefore, the velocity shift of the transit.  

On the other hand, the stellar wind may have a significant effect on the velocity shift of the transmission spectrum. {\rc If the planet is outside the Alfvén radius of the star, the ram pressure of the stellar wind dominates over the magnetic pressure of the stellar wind and the magnetic field lines in the stellar wind follow the stellar wind plasma.    

The effect of the stellar wind on the geometry of the planetary outflow has been studied in both hydrodynamic and MHD simulations. As the strength of the stellar wind is increased compared to the planetary outflow or planetary magnetic field, the stellar wind is able to confine the planetary outflow closer to the surface of the planet. On scales larger than the Hill sphere of the planet, the stellar wind will shape the outflow into a cometary tail that is being radially accelerated away from the star \citep[e.g.,][]{Matsakos2015, Khodachenko2019, McCann2019}. In essentially isothermal 3D hydrodynamic simulations including helium, \citet{Macleod2022} showed that this could result in both blue-shifted absorption during the optical transit and significant blue-shifted post-optical absorption. This is also apparent in 3D radiation-hydrodynamic simulations performed by \citet{Wang2021}. 

MHD simulations, including the radial magnetic field of the stellar wind, have also shown that the stellar wind causes planetary magnetic field lines to bend towards the night side \citep[e.g.,][]{Kislyakova2014, Carolan2021b, Khodachenko2021, Ben-Jaffel2022}. It is possible that this could result in a blue-shifted absorption signature during the transit of the planet, even if the planet had a reasonably strong magnetic field. However, it is likely that this would also cause a significant blue-shifted post-optical absorption like that seen in the hydrodynamic case. More 3D simulations are necessary to understand how degenerate observations are over the full parameter space.}

\subsubsection{Eccentricity Uncertainty}

To definitively assert that the velocity shift is due to the outflow, we must know the planet's eccentricity. Since highly irradiated planets are close to their host stars, they are generally assumed to have undergone orbital circularization. Small deviations away from a circular orbit can cause an apparent velocity shift, $\Delta v_{\text{los}}$, of:  

\begin{align}
    {\rc \Delta v_{\text{los}} \sim v_k e \text{cos}\omega}
\end{align}

where $v_k$ is Keplerian velocity, {\tworc e is the eccentricity} and $\omega$ is the argument of periastron of the planet \citep[e.g][]{Murray2010}. \citet{VanEylen2019} determined the eccentricities for a sample of Kepler planets through a combination of astroseismology and transit light-curve analysis. However, uncertainties in the eccentricity for individual planets were rarely $< 0.1$. For a planet with a typical orbital velocity of $\sim 100 \text{ km s}^{-1}$, an eccentricity uncertainty of $\sim $0.1 can lead to a velocity shift of $\sim 10 \text{ km s}^{-1}$, which is on the order of the velocity shift of due to the outflow geometry. This is also true when TTVs are used to determine eccentricities in multi-planetary systems  \citep{VanEylen2015}. 

{\tworc The detection of a secondary eclipse can be used to measure $e\text{cos}\omega$ of individual planets precisely \citep[e.g][]{Winn2010}, such that we can rule out this as a source of the observed velocity shift.} Due to their high temperatures and large size, this is often possible for hot Jupiters \citep[e.g,][]{Charbonneau2005, Baskin2013}. Previously, this has not generally been the case for sub-Neptunes. However, with JWST, detecting the secondary eclipse of some sub-Neptunes might be possible.  

\subsection{Implications for Real Systems}

Ideally, we would be able to use helium observations to constrain these planets' magnetic fields. As highlighted in the previous section, eccentricity uncertainties generally mean that we cannot make claims about the magnetic field strengths of individual planets based on these observations. However, we speculate about trends in the population.   

Table \ref{tab:He_Obs} lists spectroscopic detections of helium escape. For each planet, we quote the observed velocity shift of the absorption relative to the planet's rest frame, which is often determined by assuming a circular orbit if the planet is not known to be eccentric. These are not directly comparable to our models, as the observed transit depth is generally computed by integrating over all in-transit phases. We also estimate a critical surface magnetic field strength $B_c$ at which the flow transitions from "hydrodynamic" to "magnetically controlled" and no longer expect to see a blue-shifted transit spectrum. {\rc We approximate this value as the maximum surface magnetic field so that the outflow opens all field lines. Although, we caution that this is likely to be an overestimate. Since the outflow may be able to wrap around to the night side in the terminator region before all field lines are opened, 3D MHD simulations will be required to investigate this quantitatively. }

{\rc This critical magnetic field strength is found by requiring that the thermal pressure exceeds the magnetic pressure for $r > R_p$. We estimate the thermal pressure by i) assuming the outflow is an isothermal Parker wind with a fixed sound speed and mass loss rate or ii) assuming the outflow is in photoionization equilibrium. In both cases, the detailed procedure used to calculate the critical magnetic field is explained in Appedix~\ref{appendixB}. In Table~\ref{tab:He_Obs}, we give three critical magnetic field strength values. The first two are calculated assuming the isothermal Parker wind case, where the sound speed is $0.3 \times 10^4 \text{K} / 10^4 \text{K}$, and the mass loss rate is given by the energy-limited mass loss rate (Equation~\ref{eq:el_mlr}) with efficiency $\eta = 0.1$. The final estimate is given by the photoionization equilibrium case, where the gas temperature $~ 10^4$ K. In general, this last method gives a slightly larger value for the critical magnetic field but is the least applicable to old planets.}

To calculate the critical magnetic field strength, we needed to estimate the $F_{\text{uv}}$ received by each planet. For this, we used values given in each of the cited detection papers (see Table~\ref{tab:He_Obs}). If no estimate was present, we used estimates from \citet{Kirk2022} in which they found the XUV flux by comparing to stars of similar spectra from the MUSCLES survey \citep{MUSCLES1, Loyd2016, Youngblood2016}. {\rc We note that the reconstructed XUV flux has significant uncertainties, and our methods for calculating the thermal pressure of the outflow are only approximate. Therefore we expect that our estimates for $B_{\text{c}}$ may vary by an order of magnitude.} 

We also estimate the extent to which the stellar wind can confine the outflow. To effectively radially accelerate the planetary gas to produce a blue-shifted transit, the stellar wind must be able to penetrate close to the planetary surface. We approximately calculate the position of the bow shock produced when the stellar wind and planetary outflow collide. The bow shock is defined by the condition that the momentum fluxes of the stellar wind and planetary outflow balance. In the simple case, where outflows are spherical, we can approximate the radial distance of the bow shock from the planet $r_b$:      

\begin{align}
r_{b} = a \sqrt{\frac{\dot{M}c_s}{\dot{M}_{*}u_{*}}}
\label{eq:bow_shock_hydro}
\end{align}

\noindent where $\dot{M}_{*}$ is the mass loss rate of the star, $\dot{M}$ is the mass loss rate of the planet, $u_{*}$ is the stellar wind velocity and $a$ is the semi-major axis of the planet. To calculate $r_b$, {\rc we estimate the mass loss rate of the planet using the energy-limited mass loss rate (Equation~\ref{eq:el_mlr}) with efficiency $\eta = 0.1$} and use solar-like wind values of $\dot{M}_{*} = 10^{-14}M_{\odot}\text{ yr}^{-1}$ and $u_{*} = 150 \text{ km s}^{-1}$.  If significant absorption occurs outside this radius, then the stellar wind is likely to play an important role in determining the shape of the transit. We also note that a strong planetary magnetic field can considerably increase the bow shock distance. In this case, the bow shock distance is approximated by the point at which the stellar wind ram pressure and magnetic pressure of the planetary field balance. Therefore, we also calculate the position of the bow shock considering the effects of a 1 Gauss surface magnetic field. Here the expression for $r_b$ is given by:  

\begin{align}
r_{b} = \left(\frac{B_0^2a^2}{2\dot{M}_{*}u_{*}}\right)^\frac{1}{4}
\label{eq:bow_shock_magnetic}
\end{align}

Interestingly, ten out of the twelve systems show blue-shifted absorption. Thus, it is highly unlikely that uncertainties in planetary motion can account for all of these. Therefore, it is likely that some of these blue-shifts can be attributed to the geometry of the outflow. There are two explanations for this. The first possibility is that these planets have relatively weak surface magnetic fields (B $\lesssim 0.1$ Gauss), and day-to-night winds are responsible for the blue-shifted transit. For sub-Neptunes, this would agree with the findings of \citet{Owen2019}, which demonstrated that the suppression of atmospheric mass loss by the presence of surface magnetic fields $\gtrsim 0.3 \text{ Gauss}$ is not consistent with both the position of the radius valley and the rocky composition of planets below the valley. This is not inconsistent with dynamo-generated magnetic fields in the cores of sub-Neptunes being on the order of a few Gauss, like solar system planets, as a few percent by mass hydrogen/helium envelope can greatly increase the radius of the planet and particularly raise the upper atmosphere where atmospheric mass loss is taking place. Since magnetic field strength for a dipole field scales like $r^{-3}$, the magnetic field can be much weaker in these regions. {\rc For hot Jupiters, this conclusion would favour the argument of \citet{Griebmeier2004} that slow rotation of hot Jupiters due to tidal locking leads to magnetic fields that are orders of magnitude smaller than Jupiters. This is in opposition to the argument of \citet{Christensen2009} which predicts magnetic field strengths an order of magnitude greater than Jupiters, based on a scaling relation between magnetic field strength and energy flux.} 

The other possibility is that stellar winds are able to penetrate close to the planetary surface and radially accelerate the gas to produce the observed blue-shifts. We note that a comet-like tail of escaping helium extending out to $\sim 7$ planetary radii has been detected in the case WASP-107b \citep{Spake2021} and to a lesser extent for WASP-69b \citep{Nortmann2018}, HAT-P-18b \citep{Fu2022} and HD189733b \citep{Guilluy2020}. This may be evidence that the stellar wind plays an role in shaping the outflow of some these planets. {\rc We can set a lower limit on the extent of the outflow by assuming that the gas is optically thick to 10830 \AA~ radiation, which can compared to the bow shock radius to assess the importance of stellar wind. We call this the "minimum absorption area", and calculate its value at mid-transit (see Table~\ref{tab:He_Obs}). We stress the gas is optically thin \citep[e.g][]{Macleod2022} and therefore the actual extent can be significantly larger than this estimate. A good example of this is the transit of WASP-107b, which is found to a have a much smaller minimum absorption area $\sim 2 R_p$, than is temporally observed $\sim 7 R_p$ \citep[e.g,][]{Spake2022}}. 

In general, if the planets have weak magnetic fields, solar or moderately supersolar winds are required to set the bow shock radius close enough to the planet to disrupt the planetary outflow significantly. This is certainly feasible based on estimates of stellar wind mass rates from detections of astrospheric absorption \citep[e.g.,][]{Wood2005, Wood2021}. However, if these planets are able to host significant surface magnetic fields ($B\gtrsim 1$ Gauss), then stellar winds with mass loss rates orders of magnitudes greater than the Sun would be required. Thus, we speculate that regarding the population of close-in planets for which helium observations have been taken, there is no evidence that these planets possess magnetic field strengths strong enough to control the topology of any photoevaporative outflow.  

\begin{table*}
 \caption{Detections of helium atmospheric escape using high-resolution spectroscopy and the observed velocity shift of the outflow in the rest frame of the planet. {\rc We have provided uncertainties on the observed velocity shift, where given in the detection paper. These are generally of the order of a few km s$^{-1}$.} We estimate the maximum surface magnetic field $B_{\text{c}}$ such that all field lines are opened by the planetary outflow. {\rc The three numbers correspond to estimating strengths when we approximate the outflow to be i) an isothermal Parker wind at $10^4$ K, ii) an isothermal Parker wind at $0.3 \times 10^4$ K and iii) in photoionization-recombination equilibrium}. For the former two calculations, we estimate the mass loss rate of the planet using the energy-limited formula with an efficiency of 0.1. We also estimate the bow shock radii at which the stellar wind collides with the planetary outflow (Equation \ref{eq:bow_shock_hydro}) and the magnetosphere corresponding to a planet with a surface magnetic field of 1 Gauss (Equation \ref{eq:bow_shock_magnetic}). {\rc The minimum absorption area is a lower limit on the extent of the outflow assuming it is optically thick.}}
 \label{tab:He_Obs}
 \begin{tabular}{p{1.5cm}p{1cm}p{1.2cm}p{2.35cm}p{2.2cm}p{2.1cm}p{1.6cm}p{3.3cm}}
 \hline
Planet Name & Radius ($R_J$) & $B_{\text{c}}$ (Gauss) & Bow Shock Radius Hydrodynamic $(R_p)$ & Bow Shock Radius Magnetic $(R_p)$ & Observed Velocity Shift ($\text{km s}^{-1}$) & Minimum Absorber Area $(R_p)$ & Reference\\ 
  \hline
  GJ 1214b & 0.24 & 0.003 \newline 0.005 \newline 0.15 & 0.2 & 7.1 & $-5^{+4}_{-7}$ & 1.6 & \citet{Orell-Miquel2022}$^{(1)}$\\
  GJ 3470b & 0.36 & 0.01 \newline 0.02 \newline 0.24 & 1.2 & 10.5 & $-3.2 \pm 1.3$ & 1.9 & \citet{Palle2020}\\
  Hat-P-11b & 0.39 & 0.01\newline 0.05\newline 0.30 & 1.7 & 13.5 & $-3$ & 2.2 & \citet{Allart2018}\\
  HD 189733b & 1.13 & 0.19\newline 1.56\newline 0.54 & 1.1 & 10.3 & $-3.5 \pm 0.4$ & 1.2 & \citet{Salz2018}\\
  HD 209458b & 1.39 & 0.02\newline 0.15\newline 0.23 & 0.7 & 12.9 & $-1.8 \pm 1.3$ & 1.3 & \citet{Alonso-Floriano2019}\\
  TOI-1430.01 & 0.18 & 0.02\newline 0.05\newline 0.45 & 5.4 & 15.7 & $-4.0 \pm 1.3$ & 3.6 & \citet{ZhangM2023}\\
  TOI-1683.01 & 0.18 & 0.03\newline0.08\newline 0.53 & 3.4 & 11.3 & $-6.7 \pm 2.8$ & 3.4 & \citet{ZhangM2023}\\
  TOI-2076b & 0.22 & 0.03\newline 0.06\newline 0.46 & 5.7 & 14.9 & $-6.7 \pm 1.4$ & 3.5 & \citet{ZhangM2023}$^{(2)}$\\
  TOI-560b & 0.25 & 0.02\newline 0.04\newline 0.38 & 3.8 & 14.6 & \ \ \ $4.3 \pm 1.4$ & 2.4 & \citet{Zhang2022c}\\
  Wasp-107b & 0.92 & 0.01\newline 0.03\newline 0.23 & 2.5 & 13.9 & $-2.4 \pm 0.4$ & 2.2 & \citet{Kirk2020}\\
  Wasp-52b & 1.2 & 0.06\newline 0.43\newline 0.49 & 2.2 & 9.6 & \ \ \ $0.0 \pm 1.2$ & 1.6 & \citet{Kirk2022}\\
  Wasp-69b & 1.0 & 0.02\newline 0.13\newline 0.32 & 2.0 & 12.6 & $-3.6 \pm 0.2$ & 1.8 & \citet{Nortmann2018}\\
  \hline
 \end{tabular}
\begin{minipage}{18cm}
\vspace{0.1cm}
\vspace{0.1cm}
\rc Notes: (1) Subsequent observations by \citet{Kasper2020, Spake2022} have failed to detect escaping helium.\\ (2) \citet{Gaidos2023} also observed this planet but attributed the detection to stellar activity.     
\end{minipage}
\end{table*}

\section*{Acknowledgements}
We thank the anonymous reviewer for their comments which improved the manuscript. We are grateful to James Kirk for useful discussions. JEO is supported by a Royal Society University Research Fellowship. This project has received funding from the European Research Council (ERC) under the European Union’s Horizon 2020 research and innovation programme (Grant agreement No. 853022, PEVAP). For the purpose of open access, the authors have applied a Creative Commons Attribution (CC-BY) licence to any Author Accepted Manuscript version arising.

\section*{Data Availability}
The data underlying this article will be shared on reasonable request to the authors.



\bibliographystyle{mnras}
\bibliography{Bibliography} 



\newpage

\appendix

\section{Additional Transit Profiles} \label{appendixA}

\begin{figure}
\centering
\includegraphics[width = 0.45\textwidth]{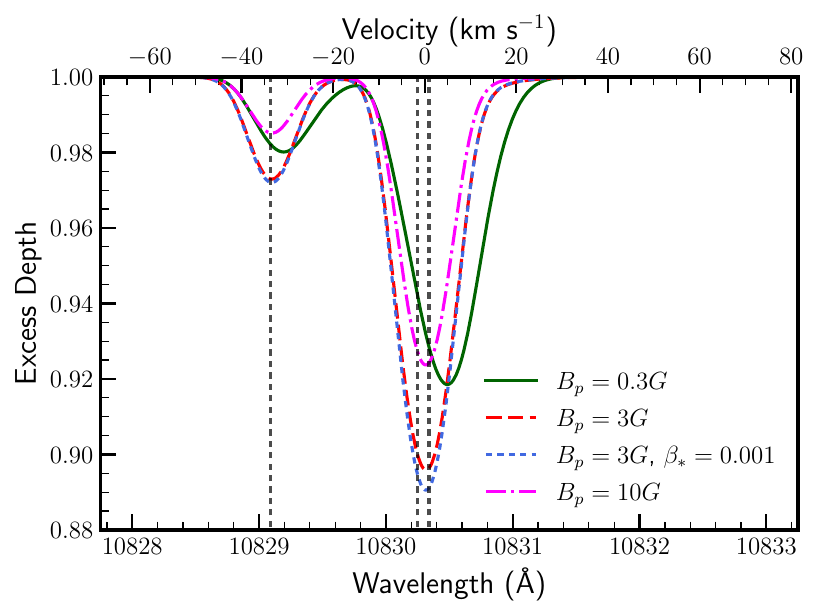}
\caption{Calculated excess absorption as a function of wavelength for different planetary magnetic field strengths where we took into account gas within 8$R_p$ of the planet. The 0.3 Gauss case now has an extended red wing due to absorption of high velocity gas far from the star. }
\label{fig:mhd_transits8Rp}
\end{figure}

Transit spectra when gas within $8R_p$ is taken into account rather than $4R_p$ are shown in Figure~\ref{fig:mhd_transits8Rp}. They are qualitatively similar to our standard case.  

\section{Solving for the critical magnetic field strength} \label{appendixB}
\subsection{Assuming the outflow is an isothermal Parker Wind}

Here, we outline the procedure to obtain the critical magnetic strength when we approximate the outflow to be an isothermal Parker wind. For a given mass loss rate and sound speed, the density and gas pressure can be written analytically in terms of the Lambert W function \citep{Cranmer2004}. The magnetic pressure along this streamline is given by:

\begin{align}
\label{eq:Pb}
P_B = \frac{B_0^2}{8 \pi}\left(\frac{R_p}{r}\right)^6
\end{align}

\begin{figure*}
\centering
\includegraphics[width = \textwidth]{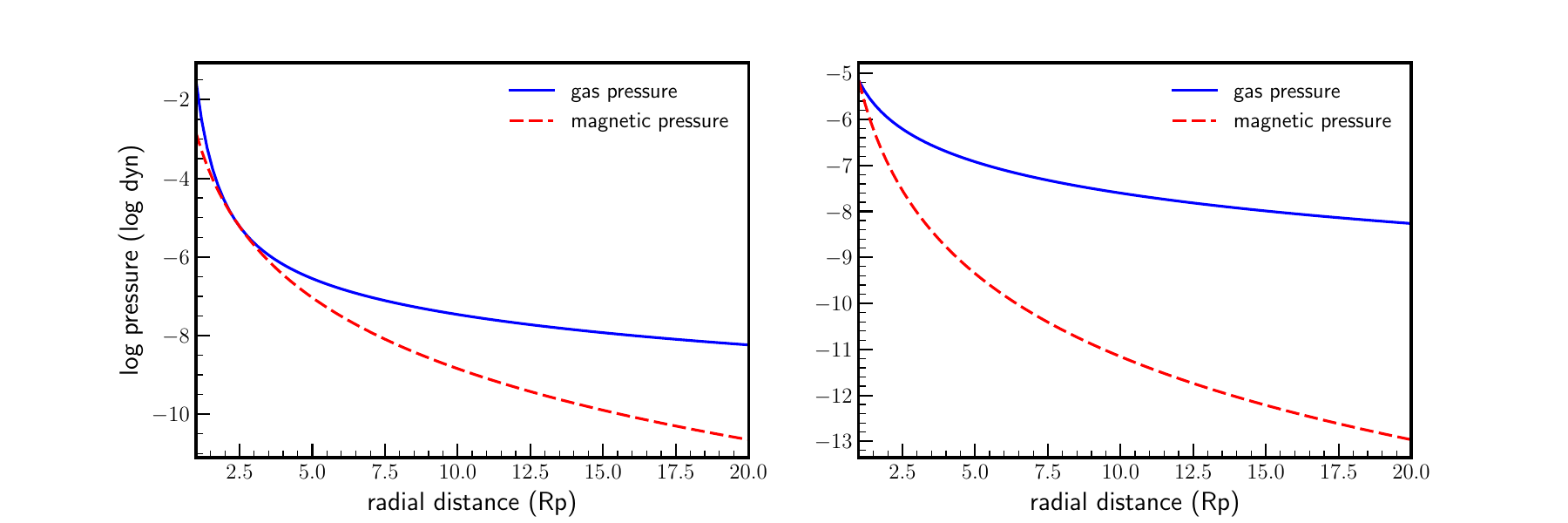}
\caption{A comparison between the magnetic pressure and gas pressure at the critical magnetic field strength for HD189733b (left) and Wasp-107b (right) assuming the outflow is an isothermal Parker wind at $10^4$ K.}
\label{fig:Bcrit}
\end{figure*}

The critical magnetic field is given by the value of the surface magnetic field such that gas pressure exceeds the magnetic pressure over for $r > R_p$ except one point where they are equal, which denote $r_B$. This point is either $R_p$ or at this point the magnetic pressure and thermal pressure are tangent. If they are tangent at this point, the thermal and magnetic pressure should satisfy:

\begin{align}
\label{eq:tangent_eq}
&P_{\text{th}}(r_B) = P_B(r_B)\\
&\frac{dP_{\text{th}}}{dr}(r_B) = \frac{dP_B}{dr}(r_B)
\end{align}
We can combine these two equations by inserting the form of the magnetic pressure (Equation~\ref{eq:Pb}) in order to get an equation just in terms of the thermal pressure. 

\begin{align}
\label{eq:th_press_eq}
\frac{dP_{\text{th}}}{dr}(r) + \frac{6P_{\text{th}}(r)}{r} = 0
\end{align}
Using the Lambert W function \citep{Cranmer2004}, the thermal pressure is given by:  

\begin{align}
&P_{\text{th}}(r) = \frac{\dot{M}c_s^2}{4\pi r^2 u(r)}\\
&u = 
\begin{cases}
-c_s^2W_0[-D(r)] \quad \ r \leq r_{c}\\
-c_s^2W_{-1}[-D(r)] \quad r > r_{c}
\end{cases}\\
&D(r) = \left(\frac{r}{r_{c}}\right)^{-4}\text{exp}\left[4\left(1 - \frac{r_c}{r}\right) - 1\right]
\end{align}
and the derivative of the thermal pressure is given by:
\begin{align}
&\frac{dP}{dr} = -\frac{\dot{M}c_s^2(2ru(r) + r^2\frac{du}{dr})}{4\pi r^4u(r)^2}\\
&\frac{du}{dr} = 
\begin{cases}
c_s\frac{\sqrt{-W_0[-D(r)]}}{2D(r)(W_0[-D(r)] + 1)} \quad \ r \leq r_{c}\\
c_s\frac{\sqrt{-W_{-1}[-D(r)]}}{2D(r)(W_{-1}[-D(r)] + 1)} \quad \ r \geq r_{c}\\
\end{cases}
\end{align}
Equation~\ref{eq:th_press_eq} can be numerically solved for $r_B$. If $r_B > R_p$, the conditions of Equation~\ref{eq:tangent_eq} are satisfied and the critical magnetic field strength is given by:  

\begin{align}
B_c = \sqrt{\frac{2\dot{M}c_s^2}{r_B^2 u(r_B)}\left(\frac{r_B}{R_p}\right)^6}
\end{align}
An example of this case is shown in the left hand panel of Figure~\ref{fig:Bcrit} (corresponding to HD 189733b). If $r_B < R_p$, there does not exist a magnetic field strength in which the conditions of Equation~\ref{eq:tangent_eq} are satisfied. Therefore we set the critical magnetic field strength to the be such that the magnetic pressure equals to gas pressure at $R_p$, which is equivalent to setting $P_{\text{th}}(R_p) = P_B(R_p)$. Therefore $B_c$ is given by: 

\begin{align}
B_c = \sqrt{\frac{2\dot{M}c_s^2}{R_p^2 u(R_p)}}
\end{align}
An example of this case is shown in the right hand panel of Figure~\ref{fig:Bcrit} (corresponding to Wasp-107b).

\subsection{Assuming the outflow is photoionization equilibrium}

For completeness, we also estimate this critical transition assuming that the flow is in photoionization equilibrium \citep[e.g][]{Owen2014, Owen2016}. We note that this is not the case for old low mass planets. In this case, we approximate that all field lines are opened if the pressure at the base of the flow is greater than the magnetic pressure. We denote this ratio as $\kappa$:  

\begin{align}
\kappa = \frac{B_0^2}{8\pi P_0}
\end{align}
where $P_0$ is the surface pressure. {\tworc The surface pressure is calculated by balancing the number of incoming photons with the number of recombinations at the ionization front, assuming the ionization front is located deep in the planet's atmosphere. Using Equation (50) from \citet{Owen2014} as our starting point}, the surface pressure can be cast in terms of the planetary properties and the incident flux.

\begin{align} 
P_0 \approx 5 \times 10^{-13} \sqrt{\frac{GM_pc_s^2F_{\text{uv}}}{R_p^2}} \text{ dyn}
\end{align}
Where $F_{\text{uv}}$ is the incident EUV flux. Therefore, the critical magnetic field strength becomes 

\begin{align}
B_c \approx 0.25 \left(\frac{M_p}{M_J}\right)^{\frac{1}{4}}\left(\frac{F_{\text{uv}}}{10^4 \text{ erg s}^{-1}}\right)^\frac{1}{4}\left(\frac{c_s}{10 \text{ km s}^{-1}}\right)^{\frac{1}{2}}\left(\frac{R_p}{R_J}\right)^{-\frac{1}{2}}
\label{eq:max_mag_field}
\end{align}


\bsp	
\label{lastpage}
\end{document}